\documentclass[aps,preprint]{revtex4}
\usepackage{epsfig}
\usepackage{amsmath}
\usepackage{epsfig}

\begin{document}
\title
{Transparency of the complex PT-symmetric  potentials for coherent injection}    
\author{Zafar Ahmed$^1$, Joseph Amal Nathan$^2$, Dona Ghosh$^3$} 
\affiliation{$~^1$Nuclear Physics Division, Bhabha Atomic Research Centre,Mumbai 400 085, India\\
$~^2$Reactor Physics Design Division, Bhabha Atomic Research Centre,Mumbai 400 085, India\\
$~^3$Department of Mathematics, Jadavpur University, Kolkata, 700032, India}
\email{1:zahmed@barc.gov.in, 2:josephan@barc.gov.in, 3: rimidonaghosh@gmail.com}
\date{\today}
\begin{abstract}
Two port s-matrix for a complex PT-symmetric potential may have uni-modular eigenvalues. If this happens for all energies, there occurs a perfect emission of waves at both ends. We call this phenomenon transparency which is distinctly different from coherent perfect absorption with or without lasing. Using the versatile PT-symmetric complex Scarf II (scattering) potential, we demonstrate analytically that the transparency can occur regardless of whether PT-symmetry is unbroken or broken or  if there are only scattering states. In these three cases, for a given value of the strength of the real part; the strength of the imaginary part $|V_2|$  of the potential lies in $(0, V_{\alpha}), (V_{\alpha}, V_{\beta})$ and  $(0,V_{\beta})$ respectively. Several other numerically solved potentials also support our findings.
\end{abstract}
\maketitle
Starting with the handedness (non-reciprocity) [1-7] of reflection probability for left and right injection of waves at a complex  potential, there have been very interesting developments in one-dimensional scattering from
complex potentials. These phenomena are spectral singularity [7,8], coherent perfect absorption without [9] and with [10] lasing, uni-modular
eigenvalues of s-matrix [11], invisibility [12], pseudo-unitarity [13], anisotropic transmission resonances [13]. These novel phenomena[7,14] have been proposed mostly based on  very strong intuition. Therefore, they  occur as a possibility  rather than a necessity  in scattering from complex non-Hermitain potentials/mediums. Although at the time of their proposals, these effects appeared to occur more generally, later they were found to have interesting limitations.

For instance for complex non-Hermitian potentials spectral singularity (SS)  means a discrete positive energy at which both the transmission and reflection probabilities become infinite. It now turns out that SS does not [15]  occur in the case of complex PT-symmetric potentials with unbroken PT-symmetry. Coherent Perfect Absorption (CPA) without lasing  was supposed to be the property of complex mediums and it has now been proved [15] that complex PT-symmetric potentials are exceptional in this regard. CPA with lasing [10] was claimed to be the property of PT-symmetric (equal gain/loss) mediums, later it has been found [11] to be a property of the domains of broken PT-symmetry.  Therefore the study and analysis  of various models of complex PT-symmetric scattering potential like 
\begin{equation}
V(x)=V_1 \phi_e(x)+i V_2 \phi_o(x), \quad \phi_{e,o}(\pm \infty)=0,
\end{equation}
in the Schr{\"o}dinger equation 
\begin{equation}
\frac{d^2 \Psi(x)}{d x^2}+\frac{2\mu}{\hbar^2}[E-V(x)]\Psi(x)=0,
\end{equation}
becomes important and valuable. In Eq. (1), $\phi_e(x)$ and $\phi_o(x)$ are real, continuous, even and odd functions (see Fig. 2), respectively. These functions may also be piece-wise continuous as the rectangular profile in Eq. (16). In all the calculations in the text and in the appendix 1, we use $2\mu=1=\hbar^2$.

For the injection of two identical waves in all respects at a potential from left and right the s-matrix connecting the out-going and incoming waves is given as [11,13]:
\begin{equation}
S= \left (\begin{array} {cc} r_{left}  & t \\  t & r_{right}  \\ \end{array} \right).
\end{equation}
Here $t$ and $r$ are transmission and reflection (complex) amplitudes
and later we use $T=|t|^2$ and $R=|r|^2$ as real transmission and reflection coefficients (probabilities). Transmission amplitude is invariant of the side (left/right) of  the incidence of the particle [1]. The complex eigenvalues of two-port $S$-matrix (6)
are  from $\det|S-s I|=0$ as
\begin{equation}
s_{\pm}={r_{left}+r_{right} \pm \sqrt{(r_{left}-r_{right})^2+4t^2} \over 2},
\end{equation}
which becomes unimodular if [11,13]
\begin{equation}
{\cal B}(k)= \left | \frac{r_{left}}{t}- \frac{r_{right}}{t} \right| \le 2.
\end{equation}
But whether this condition will be met in a some parametric domains of a given PT-symmetric potential of the type (1)  requires calculations.

The characteristic equation of this matrix is $s^2-(s_+ +s_-)s+s_+ s_-=0$. According to Caley-Hamilton theorem the matrix $S$ itself will satisfy this equation as $S^2-(s_+ +s_-)S + s_+ s_- I=0$. Further since a $2\times 2$ matrix always satisfies a quadratic, we also have $S^2-S ~\mbox{tr} (S) + det|S| I=0$. On comparing the last two equations we get  $s_+ s_-= \det S$. Then in view of the result that for PT-symmetric potentials we have $|\det|S||=1$ [15], two cases may arise. Firstly, when one of $|s_-|$ and $|s_+|$ is $<1$ and the other one $>1$. Secondly, both the eigenvalues  are uni-modular: $|s_+|=1=|s_-|$, representing perfect emission of identically (coherently) injected beam of particles at both sides of the PT-symmetric potential. 

For usual Hermitian potentials  we have $r_{left}=e^{i\alpha} r_{right}$, where $\alpha$ is a real trivial phase. Hence  both reciprocity of reflectivity and transparency are the property of Hermitian potentials. The phenomenon of transparency to be investigated here contrasts with CPA with [10] and without [9] lasing. In the coherent scattering if $|\det(S(k=k_c))|=0$  spectral singularity is found at $k=-k_c$, $E=k_c^2$ where CPA alone occurs.
If $T(E=E_*)=\infty$ and $|\det(S(E=E_*))|=0/0$ such that $|\det(S(E=E_*\pm \epsilon))|=1$ ($\epsilon$ is arbitrarily small), CPA occurs with lasing at $E=E_*$. In this case, one of the eigenvalues of s-matrix diverges and the other one is zero [10,11]. To re-emphasize, for the transparency to occur the condition that $|s_+(E)|=1=|s_-(E)|$ has to be met for all real positive energies.

We would like  to remark that even for complex PT-symmetric scattering potentials (such that, $V(\pm \infty) =0)$, the question as to whether or not this condition will be met in various domains requires computations and search. In this work, we
report that  the versatile exactly solvable Scarf II potential [2,17] is a special potential which helps in sorting out various parametric domains for transparency  explicitly and analytically. We also confirm that other numerically solved 
complex PT-symmetric potentials display similar results.

We choose $\phi_e(x)=\mbox{sech}^2x$ and $\phi_0(x)=\tanh x~ \mbox{sech}x$ in (1) to write  the complex PT-symmetric  Scarf II potential as [2,17-21]
\begin{equation}
V(x)= -V_1 \mbox{sech}^2x +i V_2 \mbox{sech} x \tanh x, \quad V_1,V_2 \in {\cal R}
\end{equation}
and propose to discuss the occurrence of transparency in two domains: (1) when $V_1>0$ and $|V_2| \le V_1+1/4$, (2) when  $|V_2| > V_1+1/4$.\\ \\
{\bf (1): $V_1>0$ and $|V_2| \le V_1+1/4$: \\ real discrete spectrum (PT-symmetry is exact/unbroken)}\\
PT-symmetry of a complex PT-symmetric  potential is known to be exact (unbroken) [20], if it has real discrete spectrum and the energy-eigenstates are also (simultaneous) eigenstates of joint operator PT: [Parity $(x\rightarrow -x)$, Time-reversal $(i \rightarrow -i)].$ For (6),
two branches of  real discrete energy eigenvalues  are given as [18,21] 
\begin{equation}
E_n=-(n-a)^2, 0\le n<a, \quad E_m=-(m-1/2-b)^2, \quad 0 \le m< b+1/2,
\end{equation}
where 
\begin{equation}
a=[\sqrt{V_1+|V_2|+1/4}+\sqrt{V_1-|V_2| +1/4}-1]/2, ~ b=[\sqrt{V_1 +|V_2|+1/4}-\sqrt{V_1 -|V_2| +1/4}]/2,
\end{equation}
are real if $|V_2| \le V_{\alpha}=V_1+1/4$ [18]. The value $V_{\alpha}$ can now be called the exceptional point (EP [22]) of the potential (6). So for $|V_2|>V_{\alpha}$ the eigenvalues are complex conjugate pair, PT-symmetry is spontaneously broken and it follows that
\begin{equation}
PT[\psi_{E_n}(x)]= \psi_{E_n^*}(x).
\end{equation}

Using the elegant analytic amenability of the Scarf II potential for reflection and transmission amplitudes [2,17], we have earlier found the simple forms of the transmission and reflection coefficients: $T(k)$ and $R(k)$ [15]. Here we need to give the simplified  expression of ${\cal B}(k)$ for this case. We find
\begin{equation} 
{\cal B}(k)= 2|\cos \pi a \sin \pi b|~\mbox{sech} \pi k \le 2, \quad k=\sqrt{E}
\end{equation}
for real values of $a$ and $b$ (5). So it follows that transparency exists in the domain where the  PT-symmetric potential has real discrete spectrum and PT-symmetry is exact (unbroken).\\ \\
{\bf (2)  $|V_2|> V_1+1/4$: \\ PT-symmetry broken}:\\
We find that the parametrization
\begin{equation}
V_1(c)=2[(c+1/2)^2-d^2], \quad |V_2(c)|=2[d^2+(c+1/2)^2],
\end{equation}
where $V_1$ could be positive so the real part of $V(x)$ (6,11) is a well (see Fig. 2(a)) or it could be negative so that the real part of $V(x)$ is a barrier (see Fig. 2(b)). In the former case there will be no real discrete spectrum for the complex PT-symmetric potential as $|V_2(c)|>V_1(c)+1/4=V_{\alpha}(c)$, instead  there will be discrete Complex-Conjugate  Pairs (CCP) of eigenvalues along with scattering states. In the latter case there will be no discrete (real or CCP) eigenvalues, the spectrum would consist only of continuous energy scattering states. Using the available [2,17] analytic forms of $t(k)$ and $r(k)$, in this case we find that
\begin{equation}
\frac{r_{left,right}}{t}=-i\left[\mp \frac{(\cos 2\pi c +\cosh 2\pi d)}{2\cosh \pi k} + \frac{(\cosh 2\pi d -\cos 2\pi c) }{2\sinh \pi k} \right],
\end{equation}
and the transmission coefficient$(T=|t|^2)$  as
\begin{equation}
T(k)=\frac {\sinh^2 \pi k \cosh^2 \pi k}{[\sin^2\pi c+ \sinh^2 \pi (d-k)]~[\sin^2\pi c+ \sinh^2 \pi (d+k)]}.
\end{equation}
Remarkably, $T(k=d)= \infty$ i.e.,  the spectral singularity (SS) occurs at $k=d$, conditionally when  $c$ is an integer. We have already conjectured [16] that SS can occur conditionally when  PT-symmetry is broken (spectrum is devoid of real discrete eigenvalues). Also see the  Appendix 1.
Inserting the expressions (12) in (5) we obtain
\begin{equation}
{\cal B}(k)= [\cos 2c \pi+ \cosh 2d \pi] ~
\mbox{sech} \pi k \le 2.
\end{equation}
The expressions (12-14) are valid for real values of $c$ and $d$. ${\cal B}(k)$ is a monotonically decreasing function of $k$, its maximum value occurs at $k=0$ and equals ${\cal B}(0)$. When $d=0$, transparency occurs. Otherwise, $\cosh 2\pi d \ge 1$ and $-1 < \cos 2\pi c < 1$, therefore the condition of transparency will  be met within the  contour in $c-d$ plane which is determined by Eq.(14) as  
\begin{equation}
d(c) \le \frac{\cosh^{-1}[2-\cos 2\pi c]}{2\pi}, \quad  d_{max}=d(1/2)= 0.2805,
\end{equation} 
and is shown in Fig.1(a). 
Since the results (12-15) are periodic function of $c$, so without a loss of generality we can assume $c\in (-1,1)$.  The curve $V_{\beta}(c)$ is obtained by substituting $d(c)$ from Eq. (15) in $V_2(c)$ in Eq. (11). The potential (6) with (11) for any pair of $(c,d)$ as allowed by the condition (15) will enjoy transparency despite breaking of PT-symmetry. In other words in terms of the Scarf II (6), for the transparency for a given value of $c$, $V_1(c)$ is determined by (11) and $d(c)$ is restricted as per the Eq. (15). For $d>0.2805$
Scarf II cannot be transparent for any value of $c$.

The potentials (1) considered here entail scattering states essentially, they may or may not possess discrete spectrum. If $V_1(c)>0$ and $|V_2|<V_{\alpha}(c)$, the transparency is observed along with  real discrete spectrum in  $V(x)$. If $V_1(c)>0$ and $V_{\alpha}(c)<|V_2|\le V_{\beta}(c)$, the transparency is observed along with  discrete spectrum consisting of CCP eigenvalues in the well (such that ${\cal R} (E_n)\le V_1(c)$). Next, if $V_1(c)<0$ and $|V_2| \le V_{\beta}$, the transparency is observed  due only to the scattering states. For example, when we choose, $c= 0.4 (-0.4), d=0.3$ in (11) (see Fig. 2(a,b)) to get real part of the Scarf II (6) as a well (barrier). However as $d>d_{max}$, $s_{\pm}$ are not uni-modular up to $E \sim 0.06$. Hence Fig. 2(c) depicts the common scenario of non-occurrence of transparency.  However, both cases possess scattering states wherein ${\cal B}(k)>2$ for low energies. Not shown here is the case of transparency when there are only scattering states. For other models of $V(x)$ (1),  we choose $\phi_e(x)=\mbox{sech}x, \phi_o(x)=\tanh x \mbox{sech}x$; $\phi_e(x)= e^{-x^2}, \phi_o(x)=xe^{-x^2}$ and so on.

Next, we choose the piecewise constant functions $\phi_e(x)= \Theta_1(x)$ and $\phi_o(x)=\Theta_2(x)$ to construct a rectangular complex PT-symmetric potential  
\begin{equation} 
V_R(x)=-V_1 \Theta_1(x)+iV_2 \Theta_2(x), \quad
\Theta_1(x)=\left\lbrace\begin{array}{lcr}
1, & &  |x|\le L\\
0, & &  |x|>L\\
\end{array}
\right.,\quad
\Theta_2(x)=\left\lbrace\begin{array}{lcr}
0, & & |x|\ge L\\
-1, & & -L < x <0\\
1, & & 0 \le x <L \\
\end{array}
\right.
\end{equation}
For other compact support models, we choose parabolic: $\phi_e(1 \le x \le1)=(1-x^2), \phi_{o,e}(|x|\ge 1)=0, \phi_o(|x|\le 1)=x(1-x^2)$,  triangular and parabolic potential profiles.

We solve the Schr{\"o}dinger equation (2) by numerical integration for (16) and for other potentials of the type (1) to extract $t(E), r_{left}(E), r_{right}(E)$. For the injection from left, we use
\begin{eqnarray}
&&\psi(x<-L)= A e^{ikx} + B e^{-ikx} \\ \nonumber
&&\psi(|x|\le L)=f(x)\\ \nonumber
&&\psi(x \ge L)=C e^{ikx}.
\end{eqnarray}
We start the integration of Eq.(2) from right $x=L$, using $\psi(L)
=C e^{ikL}, \psi^\prime(L)=C ik e^{ikL}$. Here C is any arbitrary real or non-real number. We integrate upto $x=-L$
and save $f(-L)$ to find
\begin{equation}
r_{left}(E)=e^{-2ikL} \frac{ikf(-L)-f'(-L)}{ikf(-L)+f'(-L)}, \quad t_{left}(E)= e^{-ikL} \frac{2ik}{ikf(-L)+f'(-L)}.
\end{equation}
For the injection from right, we have
\begin{eqnarray}
&&\psi(x>L)= B e^{ikx} + A e^{-ikx} \\ \nonumber
&&\psi(|x|\le L)=g(x)\\ \nonumber
&&\psi(x <-L)=C e^{-ikx}.
\end{eqnarray}
This time we integrate Eq. (2) from $x=-L$, using $\psi(-L)= C e^{ikL}, \psi'(-L)=-C ik e^{ikL}$ and save $g(L)$.
So for the  injection from right, we get:\\
\begin{equation}
r_{left}(E)=e^{-2ikL} \frac{ikg(L)+g'(L)}{ikg(L)-g'(L)}, \quad t_{left}(E)= e^{-ikL} \frac{2ik}{ikg(L)-g'(L)}.
\end{equation}
Using this numerical procedure, we study the complex PT-symmetric potential (16) by fixing $V_1=5,a=2$  and varying $V_2$. We find that  the scattering coefficients $R_{left} \ne R_{right}$ and $T_{left}=T_{right}$ [1]. When $|V_2|\le 2.19$ in (16), these  coefficients  commonly show two poles (spikes) indicating the existence of two real discrete bound states for $E<0$. In Fig. 3, we present only  $R_{left}(E)$  to show  two poles for two closely lying bound states at negative energies. The PT-symmetry breaks down spontaneously when $|V_2|=2.20$ and poles at real discrete energies disappear to give way to a single maximum. In this case the real eigenvalue pairs first coalesce when
$c=b$ and convert to complex conjugate pairs of energy eigenvalues when $b^2 < c^2 $ (e.g., $a\pm\sqrt{b^2-c^2} \rightarrow a\pm i \sqrt{c^2-b^2}$). Thus, for this case of (16) $|V_2|=V_{\alpha}$ is the exceptional point.

In figure 4, notice that this potential (16) sustains transparency for $|V_2|=2.20, 2.30$ and ${\cal B}(k) \le 2$. However, for $|V_2|=2.40=V_{\beta}$ transparency can not be observed as $s_{\pm}$ are not uni-modular up to $E\sim 0.35$, also we get ${\cal B}(k) >2$. Not shown here are the cases when the real part of $V(x)$ is a barrier having only scattering states, yet it displays transparency. One such case is when $V_1=-0.70,|V_2|=0.10$ in (16).

We find that in the cases when the complex PT symmetric potential is of finite support  there exists an energy say $E_s$ at which an interesting transition takes place: if $|s_+(E<E_s)| > |s_-(E<E_s)|$ then  $|s_+(E>E_s)| < |s_-(E>E_s)|$. See the vertical line in the loop in Fig. 4(c). 

Apart from the potentials (6, 16) for which we present results in Figs. (1-3), we have also investigated the  potentials of the type (1), using the continuous Gaussian, \mbox{sech}, Lorentzian, triangular and parabolic profiles. The features presented by the rectangular potential (20) are also displayed by several other numerically solved potentials wherein we use parabolic, triangular, Gaussian profiles of finite support. However, in case of long ranged potentials like (un-truncated) Gaussian when transparency does not occur we do not find the said cross-over of $|s_{\pm}|$. Instead, we get the scenario (see Fig. 2(c)) like that of the long ranged Scarf II potential.

We present our conclusions based on our exact analytic expressions
obtained above for Scarf II and numerical computations of various potentials of the type (1).

\noindent
$\bullet$ Complex PT-symmetric potentials (1) share yet another 
feature common with the Hermitian potentials. This feature is transparency of the potential for coherent injection.
\\
$\bullet$ Complex PT-symmetric potentials (1) with real part as a well, have two critical values for the strength parameter of the imaginary part of the potential, they are $V_{\alpha}$ and $V_{\beta}$. The former is the exceptional point of the  non-Hermitian potential below which the potential has real discrete spectrum and PT-symmetry is unbroken. We find that above $V_{\alpha}$ there exists another exceptional value  $V_{\beta}$ upto which transparency (uni-modularity of $s_{\pm}$ at any energy) is observed. This implies that unbroken PT-symmetry is only {\it sufficient} but not {\it necessary} for transparency. The special and versatile complex Scarf II demonstrates these features explicitly and analytically (also see Fig. 1).\\
$\bullet$ The complex PT-symmetric scattering potentials (1) having real part as a barrier ($V_1\phi_e(x)>0)$ have only scattering states (continuum of positive energies) can also display transparency when $|V_2|< V_{\beta}$
\\
$\bullet$ The complex PT-symmetric potentials (2) of compact support (16) give rise to a critical value of energy, $E_s$, about which  the values $|s_{\pm}|\ne 1$  display a cross-over transition [see fig. 4(c)]. 
It will be important to check lest at $E=E_s$ two Riemann sheets are crossed as in the definitions of $|s_{\pm}|$ in Eq. (5), we have complex quantities under the square-root sign.\\
$\bullet$ The parametric domain of transparency in a complex PT-symmetric scattering potential is
devoid of spectral singularity. See transmission coefficients (13) (for $d$=0),  Eq. 22 (of Ref. [15]) and Appendix A. We conjecture yet again that the spectral singularity is not found in the domain with unbroken PT-symmetry. \\
$\bullet$ The simple analytic expressions in Eqs. (10,12-15) presented here are new.

The  sameness of complex PT-symmetric potential and real Hermitian potentials with regard to the transparency needs to be taken cautiously that in the former case only some parametric domains becomes transparent for coherent injection/incidence. We have called, the uni-modularity of the eigenvalues of s-matrix  at any energy of a complex PT-symmetric potential, transparency and investigated it in detail. We hope that the present work will generate further interest and investigations in the transparency of complex PT-symmetric mediums for coherent injection.
\section*{\Large{Appendix 1}}
\renewcommand{\theequation}{A-\arabic{equation}}
\setcounter{equation}{0}
\noindent
{\bf Spectral singularity and bound states seem to be mutually exclusive in a complex PT-symmetric potential}\\
The scattering from  the complex PT-symmetric potential
\begin{equation}
V(x)=(V_1+iV_2)\delta(x+a)+(V_1-iV_2)\delta(x-a), \quad V_1,V_2 \in {\cal R}
\end{equation}
can help seeing yet again that spectral singularity and real  discrete energy bound states are mutually exclusive. Next by writing 
the solutiion of Schr{\"o}dinger equation (2) as $\psi(x< -a) =A e^{ikx}+ B e^{-ikx}, \psi(-a<x<a)=C e^{ikx} +D e^{-ikx}, \psi(x) =F e^{ikx}$y  matching the wavefunction and mismatching their derivative (due to Delta functions (A-1)) at $x=\pm a$ we get  reflection $r(k)=B/A$ and the transmission $t(k)=F/A$ amplitudes as 
\begin{equation}
r(k)= \frac{i e^{-2ika}[(2kV_2-V_1^2-V_2^2)\sin 2ka -2kV_1\cos 2ka]}{2k^2\cos 2ka +2kV_1 \sin 2ka +i[2kV_1 \cos 2ka + (V_1^2+V_2^2-2k^2) \sin 2ka]}
\end{equation}
\begin{equation}
t(k)=\frac{-2k^2}{2k^2\cos 2ka +2kV_1 \sin 2ka +i[2kV_1 \cos 2ka + (V_1^2+V_2^2-2k^2) \sin 2ka]}
\end{equation}
reflection and transmission coefficients are found as $T=|t|^2$ and $R=|r|^2$, respectively.

In Ref.[23] the parametric PT-symmetric  regimes of $V_1,V_2$ for the bound states of (A-1) have been worked out well, it is found that there
are at most two real bound states. In Ref. [24] the condition and location of
spectral singularity for PT-symmetric domains is found to be only of two types: (1) The SS is found at $E=E_*=\frac{\pi^2(2n+1)^2}{16a^2}$
when $V_1=0$ and $|V_2|=\frac{\pi(2n+1)}{2\sqrt{2}a}, n=0,1,2,3$ [24].
(2) The SS is found at $E=E_*= [V_2^2-V_1^2]/2$ when the condition that 
$2V_1+\sqrt{2[V_2^2-V_1^2]}~ \cot[a \sqrt{V_2^2-V_1^2}]=0$ [24] is satisfied.

We use the previously suggested [23,24] parameters $V_1$ and $V_2$  for bound states [23] and spectral singularity [24] for PT-symmetric cases to calculate $T(E)$ (A-3) for both negative and positive energies to find that the poles exists either for negative energies or for positive energies. The
former are bound states (see cases 1-3 in Table 1) there are the negative energy poles $(E_0, E_1)$ consistent with [23] and the latter ones are spectral singularities (see $E_*$ in the cases 5-12 in Table 1) consistent with [24]. In these cases we find a single pole (large spike) in $T(E)$ at $E=E_*>0$, but no pole or spike at a negative energy such as displayed in Fig. 3 for the potential (16). The Table 1,  can be seen to testify the mutual exclusiveness of spectral singularity and bound states in a  complex PT-symmetric potential. Nevertheless, complex PT-symmetric Scarf II Eq. (6) displays this analytically in a simple way (see the domain (1) in the text below Eq. (6)).
\begin{table}[ht]
\caption{Both negative (bound states: BS) and positive (spectral singularity: SS) energy poles in $T(E)$ (A-3) for various values of $V_1,V_2$.  Notice that BS and SS do not co-exist. The sign of dash denotes an absence of the corresponding state.}
\begin{ruledtabular}
\begin{tabular}{|c||c||c||c||c|c|}
\hline
Case &$~~~V=V_1+|V_2|i~~~$  & Ref. &BS: $E_0$ & BS: $E_1$ & SS: $E_*$\\
\hline	
1&-2 + 0 i & [23] & -1.2295 & -0.6349 & $\--$ \\ \hline 
2 &$-2+ 0.1i$ & [23] & -1.2081 & -0.6513 & $\--$ \\  \hline
3 & $-2+0.2i$ & [23] & -1.1346 & -0.7105 & $\--$ \\ \hline
4 & $-2+0.275 i$ & $\--$ & -0.9932 & -0.8348 & $\--$ \\ \hline
5 & $-2+3.4160i$ & [24] & $\--$ & $\--$ & 3.8364 \\ \hline
6 & $-2+5.5726i$ & [24] & $\--$ & $\--$ & 13.5274 \\ \hline
7 & $+2+2.7025i$ & [24] & $\--$ & $\--$ & 1.6517\\ \hline
8 & $+2+4.2816i$ & [24] & $\--$ & $\--$ & 7.1664\\ \hline
9 & $0+3.3321i$ & [24] & $\--$ & $\--$ & 5.5516\\ \hline
10 & $0+5.5536i$ & [24] & $\--$ & $\--$ & 7.5508\\ \hline
11 & $-5+5.5735i$ & [24] & $\--$ & $\--$ & 3.0289\\ \hline
12 & $-10+15.3099i$ & [24] & $\--$ & $\--$ & 67.1959 \\ \hline
13 & $+5+6.4491i$ & [24] & $\--$ & $\--$ & 8.2960 \\ \hline
14 & $+10+ 11.8640i$ &[24]& $\---$ & $\--$ & 20.2583 \\ \hline
\end{tabular}
\end{ruledtabular}
\end{table}

\section*{Acknowledgement} We would like to thank the anonymous Referee for his valuable questions and comments. DG would like to thank Prof. Subenoy Chakraborty (Dept. of Mathematics, Jadavpur University, Kolkata) for his support and interest in this work.

\begin{figure}[H]
\centering
\includegraphics[width=7 cm,height=5 cm]{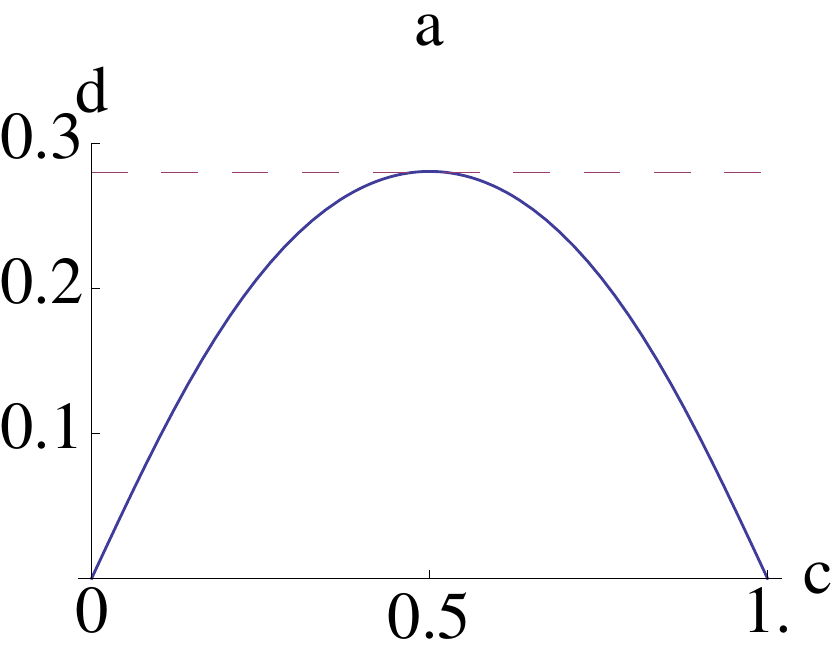}
\hskip .5 cm
\includegraphics[width=7 cm,height=5 cm]{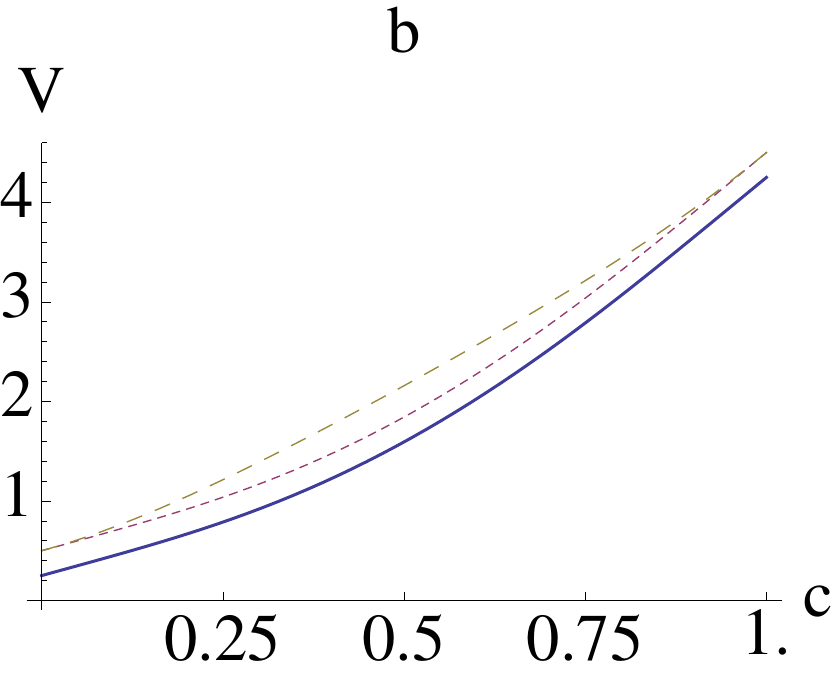}
\caption{(a): The plot of $d(c)$ (15). This curve marks the boundary above which ($d_{max}=0.28$, see the dashed line) $s_{\pm}$ will not be uni-modular for all energies of injection/incidence. (b):
The solid line denotes $V_1(c)$, the  short-dashed curve denotes $V_{\alpha}(c)=V_1(c)+ 0.25$ and  the long-dashed curve denotes $V_{\beta}$. These values are such that $V_1(c) < V_{\alpha}(c) < V_{\beta}(c)$. Given $c$ , $V_1(c)$ gets fixed then if $V_{\alpha}(c)<|V_2|<V_{\beta}(c)$, transparency is observed without
the existence  of real discrete spectrum. When $|V_2|<V_{\alpha}$, transparency is observed along with the existence of real discrete spectrum in $V(x)$.}
\end{figure}
\begin{figure}[H]
\centering
\includegraphics[width=5 cm,height=7. cm]{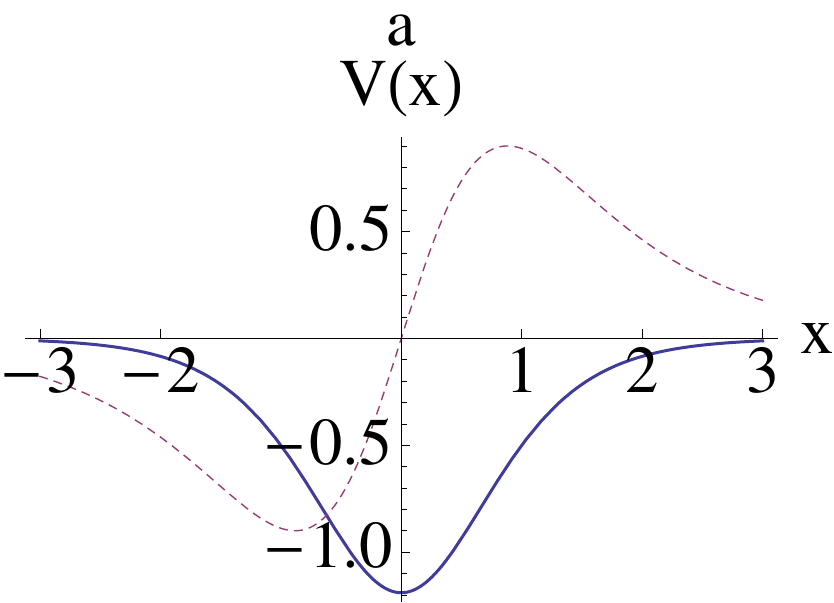}
\hskip .5 cm
\includegraphics[width=5 cm,height=7. cm]{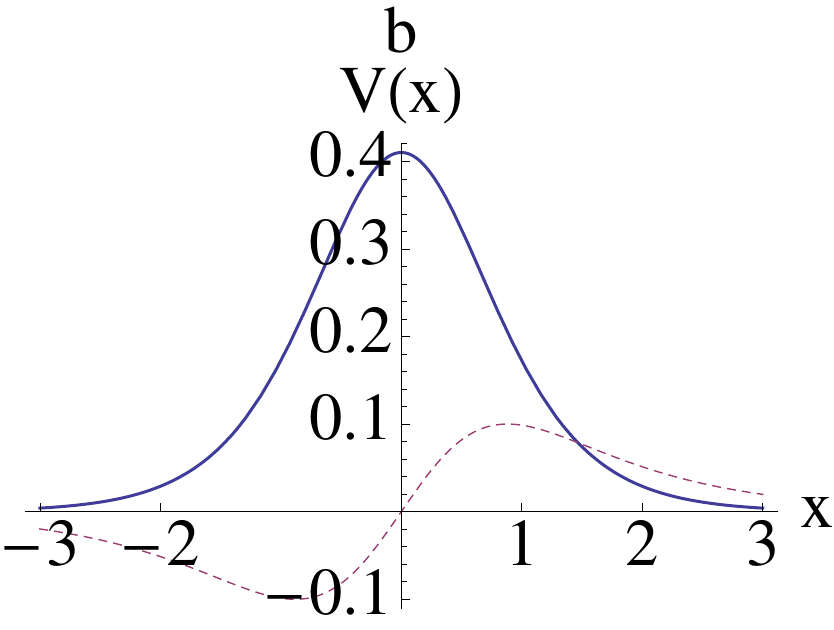}
\hskip .5 cm
\includegraphics[width=5 cm,height=7. cm]{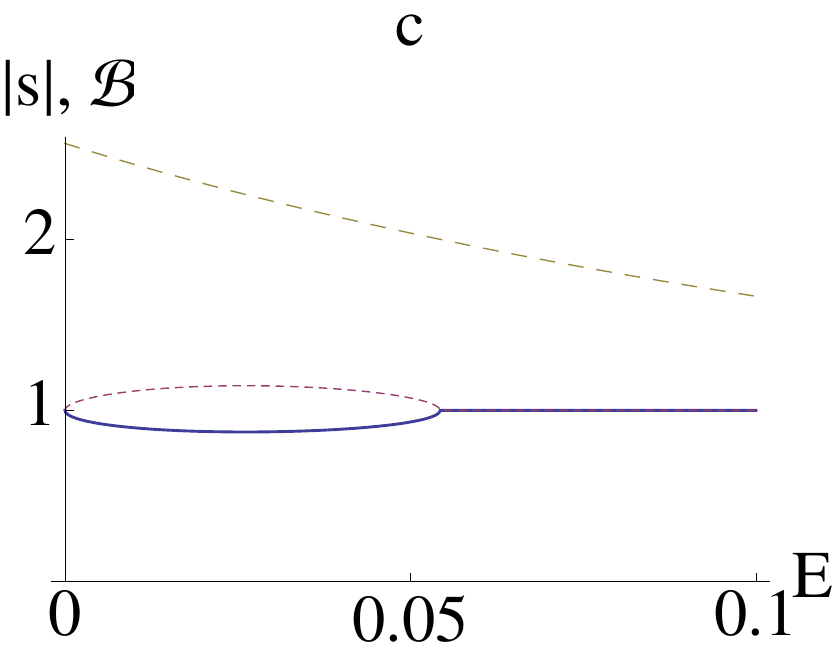}
\caption{The depiction of non-occurrence of transparency. Scarf II potentials (a): $c=0.4$ and $d=0.3$  and  (b): $c=-0.4, d=0.3$. Notice that $d>d_{max}=0.28$ (see Eq. (15)). In (a), the real part is a well and in (b) it is a barrier. In both the cases commonly and identically, the eigenvalues of s-matrix i.e., $s_{\pm}$ are not uni-modular for all energies. In (c), see the loop formed by solid and short-dashed lines for  energies up to $\sim 0.06$. For higher energies they are uni-modular. Notice that ${\cal B}(0)$ value goes above the critical value 2 at lower values of energy.} 
\end{figure}
\begin{figure}[H]
\centering
\includegraphics[width=5 cm,height=7 cm]{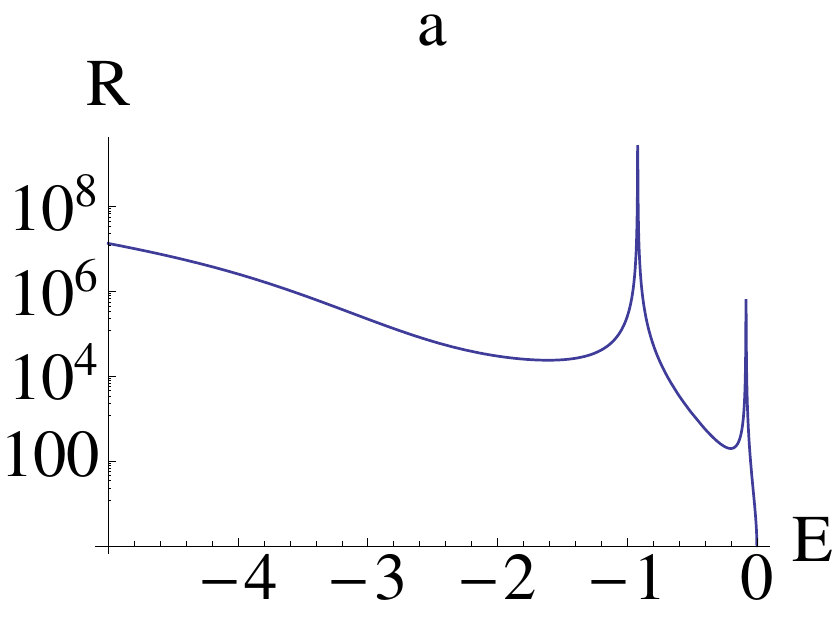}
\hskip .5 cm
\includegraphics[width=5 cm,height=7 cm]{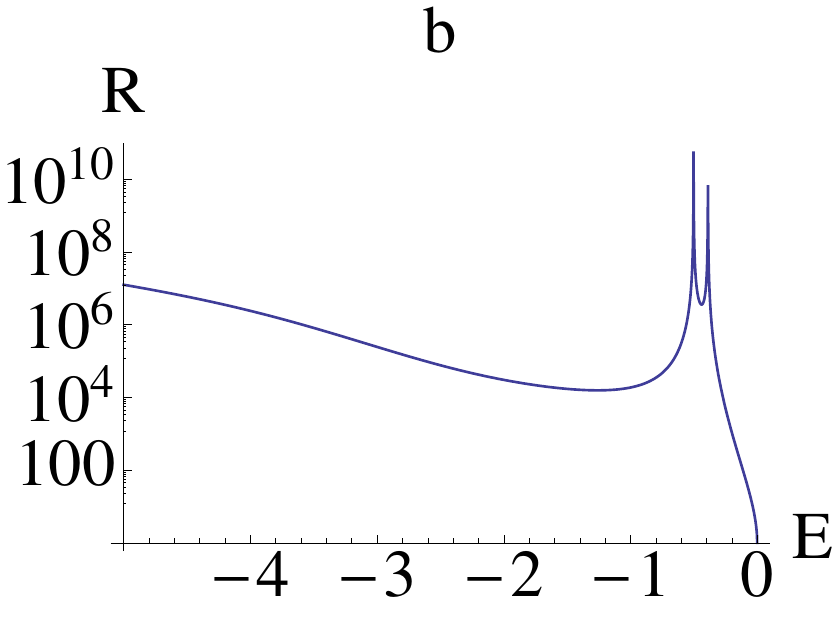}
\hskip .5 cm
\includegraphics[width=5 cm,height=7 cm]{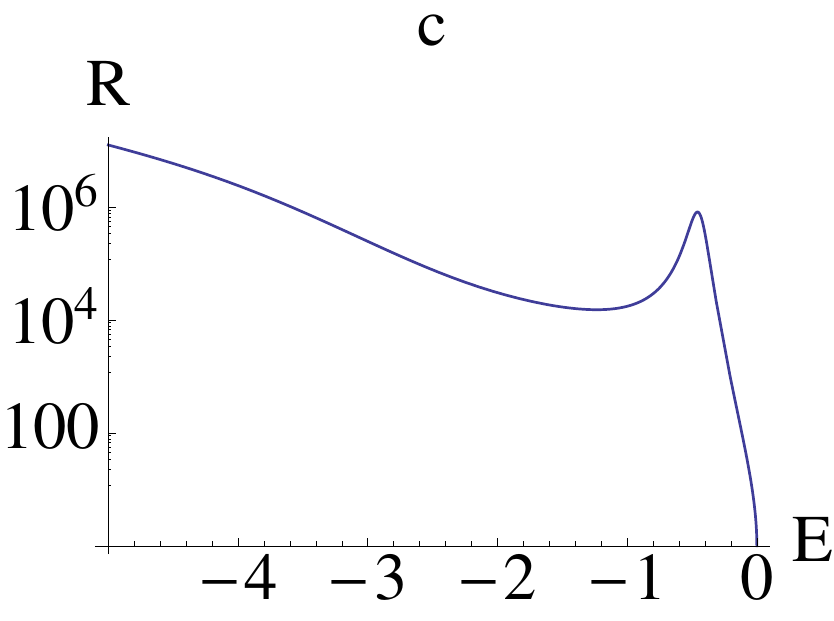}
\caption{For rectangular potential (16) with $V_1=5,a=2$, $R(E)$ for left injection/incidence is plotted to show the occurrence of two bound levels (poles) at negative energies: (a) when $V_2=2.00$. In  (b) when $V_2= 2.19$ the real eigenvalues are very close by. In (c) for $V_2=2.20$  these two levels coalesce to become complex conjugate pairs and PT-symmetry is spontaneously broken, the poles disappear to give rise to a single maximum. $V_2=2.20$ is the exceptional point of this non-Hermitian potential}
\end{figure}
\begin{figure}[H]
\centering
\includegraphics[width=5 cm,height=7 cm]{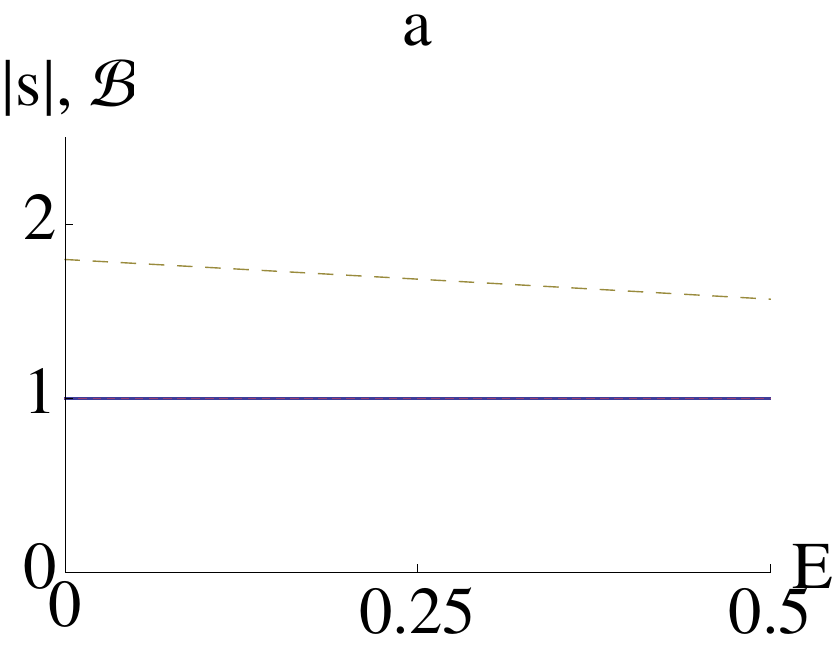}
\hskip .5 cm
\includegraphics[width=5 cm,height=7 cm]{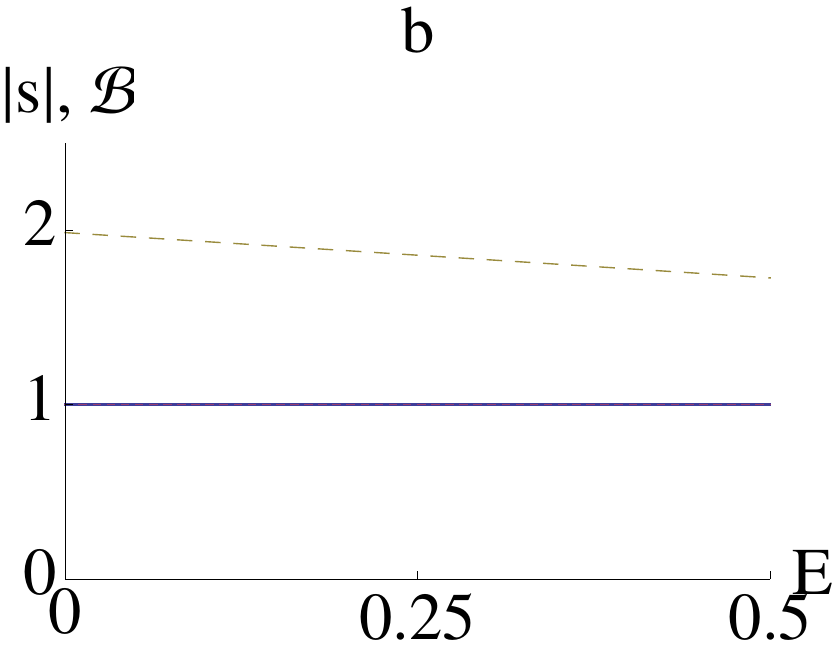}
\hskip .5 cm
\includegraphics[width=5 cm,height=7 cm]{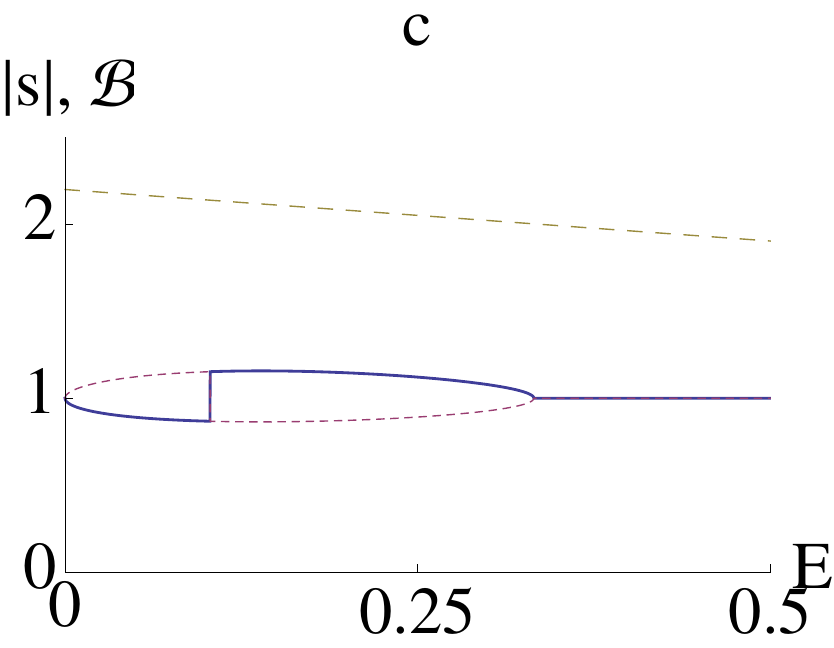}
\caption{$s_{\pm}$ for the same potential as in Fig. 3. (a): $|V_2|=2.20$  and (b): $V_2= 2.30$), the transparency is sustained. However, in (c): $V_2=2.40$  the transparency is not observed as $s_{\pm}$ do not remain uni-modular at lower energies of injection/incidence (see a loop of short dashed lines and solid line in (c)). Also see the vertical line in the loop showing the discontinuous cross-over of $|s_{\pm}|$ at $E \sim 0.12$. Also see that ${\cal B}(k)$ (long-dashed line) remain $\le 2$ in (a,b) but in (c) it is $>2$.}
\end{figure}

\end{document}